\documentclass[10pt]{article}

\usepackage{amsmath}
\usepackage{amssymb}

\usepackage{graphicx}

\usepackage{cite}

\usepackage{color} 

\usepackage{setspace} 
\usepackage{lineno}
\linenumbers

\usepackage[labelfont=bf,labelsep=period,justification=raggedright]{caption}

\makeatletter
\renewcommand{\@biblabel}[1]{\quad#1.}
\makeatother

\date{}

\begin{document}

\begin{flushleft}
{\Large
\textbf{Inferring evolutionary histories of pathway regulation from transcriptional profiling data}
}
\\
Joshua G. Schraiber$^{1}$, 
Yulia Mostovoy$^{2}$, 
Tiffany Y. Hsu$^{2, 3}$
Rachel B. Brem$^{2, \ast}$
\\
\bf{1} Department of Integrative Biology, University of California, Berkeley, CA, USA
\\
\bf{2} Department of Molecular and Cellular Biology, University of California, Berkeley, CA, USA
\\
\bf{3} Present Address: Graduate Program in Biological and Biomedical Sciences, Harvard Medical School, Boston, MA, USA
\\
$\ast$ E-mail: Corresponding rbrem@berkeley.edu
\end{flushleft}

\section*{Abstract}

One of the outstanding challenges in comparative genomics is to
interpret the evolutionary importance of regulatory variation between
species.  Rigorous molecular evolution-based methods to infer evidence for
natural selection from expression data are at a premium in the field,
and to date, phylogenetic approaches have not been well-suited to
address the question in the small sets of taxa profiled in standard
surveys of gene expression. We have developed a strategy to infer
evolutionary histories from expression profiles by analyzing suites of
genes of common function. In a manner conceptually similar to
molecular evolution models in which the evolutionary rates of DNA
sequence at multiple loci follow a gamma distribution, we modeled
expression of the genes of an \emph{a priori}-defined pathway with
rates drawn from an inverse gamma distribution. We then developed a
fitting strategy to infer the parameters of this distribution from
expression measurements, and to identify gene groups whose expression
patterns were consistent with evolutionary constraint or rapid
evolution in particular species.  Simulations confirmed the power and
accuracy of our inference method.  As an experimental testbed for our
approach, we generated and analyzed transcriptional profiles of four
\emph{Saccharomyces} yeasts.  The results revealed pathways with
signatures of constrained and accelerated regulatory evolution in
individual yeasts and across the phylogeny, highlighting the
prevalence of pathway-level expression change during the divergence of
yeast species. We anticipate that our pathway-based phylogenetic
approach will be of broad utility in the search to understand the
evolutionary relevance of regulatory change.

\section*{Author Summary}

Comparative transcriptomic studies routinely identify thousands of
genes differentially expressed between species.  The central question
in the field is whether and how such regulatory changes have been the
product of natural selection.  Can the signal of evolutionarily
relevant expression divergence be detected amid the noise of changes
resulting from genetic drift?  Our work develops a theory of gene
expression variation among a suite of genes that function together.
We derive a formalism that relates empirical observations of
expression of pathway genes in divergent species to the underlying
strength of natural selection on expression output.  We show that
fitting this type of model to simulated data accurately recapitulates
the parameters used to generate the simulation.  We then make
experimental measurements of gene expression in a panel of
single-celled eukaryotic yeast species.  To these data we apply our
inference method, and identify pathways with striking evidence for
accelerated or constrained regulatory evolution, in particular species
and across the phylogeny.  Our method provides a key advance over
previous approaches in that it maximizes the power of rigorous
molecular-evolution analysis of regulatory variation even when data
are relatively sparse.  As such, the theory and tools we have
developed will likely find broad application in the field of
comparative genomics.

\section*{Introduction}

Comparative studies of gene expression across species routinely detect
regulatory variation at thousands of loci \cite{romero2012comparative}.
Whether and how these expression changes are of evolutionary relevance
has become a central question in the field.  In landmark cases,
experimental dissection of model phenotypes has revealed evidence for
adaptive regulatory change at individual genes
\cite{rebeiz2009stepwise, chan2010adaptive,jones2012genomic,
  ishii2013oslg1}.  These findings have motivated
hypothesis-generating, genome-scale searches for signatures of natural
selection on gene regulation.  In addition to molecular-evolution
analyses of regulatory sequence \cite{torgerson2009evolutionary,
  he2011does, shibata2012extensive, gronau2013inference}, phylogenetic
methods have been developed to infer evidence for non-neutral
evolutionary change from measurements of gene expression
\cite{chaix2008evolution,
  bedford2009optimization,brawand2011evolution}.  Two classic models
of continuous character evolution have been used for the latter
purpose: Brownian motion models, which can specify lineage-specific
rates of evolution on a phylogenetic tree \cite{lande1976natural,
  felsenstein1988phylogenies, o2006testing, eastman2011novel} and have
been used to model the neutral evolution of gene expression
\cite{oakley2005comparative, bedford2009optimization}, and the
Ornstein-Uhlenbeck model, which by describing lineage-specific forces
of drift and stabilizing selection \cite{lande1976natural,
  hansen1997stabilizing, butler2004phylogenetic} can be used to test
for evolutionary constraint on gene expression
\cite{bedford2009optimization, brawand2011evolution}.  To date,
phylogenetic approaches have had relatively modest power to infer
lineage-specific rates or selective optima of gene expression levels.
This limitation is due in part to the sparse species coverage typical
of transcriptomic surveys, in contrast to studies of organismal traits
where observations in hundreds of species can be made to maximize the
power of phylogenetic inference \cite{ane2008analysis,
  harmon2010early, boettiger2012your}.

As a complement to model-based phylogenetic methods, more empirical
approaches have also been proposed that detect expression patterns
suggestive of non-neutral evolution \cite{blekhman2008gene,
  bullard2010polygenic,fraser2010evidence}.  We previously developed a
paradigm to detect species changes in selective pressure on the
regulation of a pathway, or suite of genes of common function, in the
case where multiple independent variants drive expression of pathway
genes in the same direction \cite{bullard2010polygenic,
  martin2012evolution}.  Broadly, pathway-level analyses have the
potential to uncover evidence for changes in selective pressure on a
gene group in the aggregate, when the signal at any one gene may be
too weak to emerge from genome-scale scans.  However, the currently
available tests for directional regulatory evolution are not well
suited to cases in which some components of a pathway are activated,
and others are down-regulated, in response to selection.

In this work, we set out to combine the rigor of phylogenetic methods
to reconstruct histories of continuous-character evolution with the
power of pathway-level analyses of regulatory change.  We reasoned
that an integration of these two families of methods could be used to
detect cases of pathway regulatory evolution from gene expression
data, without assuming a directional model.  To this end,
we aimed to develop a phylogenetic model of pathway regulatory change
that accounted for differences in evolutionary rate between the
individual genes of a pathway.  We sought to use this model to uncover
gene groups whose regulation has undergone accelerated evolution or
been subject to evolutionary constraint, over and above the degree
expected by drift during species divergence as estimated from genome
sequence.  As an experimental testbed for our inference strategy, we
used the \emph{Saccharomyces} yeasts.  These microbial eukaryotes span
an estimated 20 million years of divergence and have available
well-established orthologous gene calls \cite{scannell2011awesome},
and yeast pathways are well-annotated based on decades of
characterization of the model organism \emph{S. cerevisiae}.  We
generated a comparative transcriptomic data set across Saccharomycetes
by RNA-seq, and we used the data to search for cases of pathway
regulatory change.
\section*{Results}
\subsection*{Modeling the rates of regulatory evolution across the genes of a pathway}

The Brownian-motion model of expression of a gene predicts a
multivariate normal distribution of observed expression levels in the
species at the tips of a phylogenetic tree. The variance-covariance
matrix of this multivariate normal distribution reflects both the
relatedness of the species and the rate of regulatory evolution along
each branch of the tree.  We sought to apply this model to interpret
expression changes in a pre-defined set of genes of common function,
which we term a pathway.  Our goal was to test for accelerated or
constrained regulatory variation in a pathway relative to the
expectation from DNA sequence divergence, as specified by a genome
tree.  To avoid the potential for over-parameterization if the rate of
each gene in a pathway were fit separately, we instead developed a
formalism, detailed in Methods, to model regulatory evolution in the
pathway using a parametric distribution of evolutionary rates across
the genes.  This strategy parallels well-established models of the
rate of DNA sequence evolution across different sites in a locus or
genome \cite{yang1996among}.  Briefly, we assumed that each gene in
the pathway draws its rate of evolution from an inverse gamma
distribution, and we derived the relationship between the parameters
of this distribution and the likelihood of expression observations at
the tips of the tree.  For each gene, we modeled the contrasts of the
expression level in each species relative to an arbitrary species used
as a reference, to eliminate the need to estimate the ancestral
expression level.  A further normalization step, recentering the
distribution of expression across pathway genes in each species to a
mean of $0$, corrected for the effects of coherent regulatory
divergence due to drift.  This formalism enabled a maximum-likelihood
fit of the parameters describing the pathway expression distribution,
given empirical expression data, and could accommodate models of
lineage-specific regulatory evolution, in which a particular subtree
was described by distinct evolutionary rate parameters relative to the
rest of the phylogeny.  As a point of comparison, we additionally made
use of an Ornstein-Uhlenbeck (OU) model \cite{butler2004phylogenetic}:
here the rate of regulatory evolution of each gene in a pathway,
across the entire phylogeny, was drawn from an inverse-gamma
distribution, and all genes of the pathway were subject to the same
degree of stabilizing selection, again across the entire tree.

Our ultimate application of the method given a set of expression data
was to enumerate all possible Brownian motion models in which pathway
expression evolved at a distinct rate along the lineages of a subtree
relative to the rest of the phylogeny, and for each such model, apply
our fitting strategy and tabulate the likelihood of the data under the
best-fit parameter set.  To compare these likelihoods and the
analogous likelihood from the best-fit OU model of universal
constraint, we applied a standard Akaike information criterion (AIC)
\cite{ harmon2010early, gong2012evolution, slater2012integrating} to
identify strongly supported models.

\subsection*{Simulation testing of inference of pathway regulatory evolution}

As an initial test of our approach, we sought to assess the
performance of our phylogenetic inference scheme in the ideal case in
which rates of regulatory evolution of the genes of a pathway were
simulated from, and thus conformed to, the models of our theoretical
treatment.  In keeping with our experimental application below which
used a comparison of \emph{Saccharomyces} yeast species as a testbed,
we developed a simulation scheme using a molecular clock-calibrated
\emph{Saccharomyces} phylogeny \cite{scannell2011awesome} (see Figure
1a inset).  We simulated the expression of a multi-gene pathway in
which rates of evolution of the member genes were drawn from an
inverse gamma distribution. With the simulated expression data in hand
from a given generating model, we fit an OU model, an equal-rates
model, and models of evolutionary rate shifts in each subtree in
turn. 

Figure 1 shows the results of inferring the mode and rate of evolution
from data simulated under a model of accelerated regulatory change on
the branch leading to \emph{S. paradoxus}, and similar results can be
seen in Figures S1 through S5 for other rate shift models.
As expected, for very small gene groups, inference efforts did not
achieve high power or recapitulate model parameters (Figure 1a,
leftmost data point; Figure 1b, leftmost point in each cluster),
reflecting the challenges of the phylogenetic approach when applied on
a gene-by-gene basis to relatively sparse trees like the
\emph{Saccharomyces} species set. By contrast, for pathways of ten
genes or more, we observed strong AIC support for the true generating
model in cases of lineage-specific regulatory evolution, approaching
AIC weights of 100\% for the correct model if a pathway contained more
than 50 genes (Figure 1a, Figure S1 and panel a of
Figures S2-S5).  In these simulations our method also
inferred the correct magnitudes of lineage-specific shifts with high
confidence, for all but the smallest pathways (Figure 1b and panel b
of Figures S2-S5). Likewise, when applied to simulated
expression data generated under models of phylogeny-wide constraint,
our method successfully identified OU as the correct model (Figure
2a), though with biased estimates of the magnitude of the constraint
parameter when the latter was large (Figure 2b), likely due to a lack
of identifiability with the inverse-gamma rate parameter
(Figure S6).

We also sought to evaluate the robustness of our method to violations
of the underlying model. To explore the effect of our assumption of
independence between genes, we simulated a pathway in which expression
of the individual genes was coupled to one another and evolving under
an equal-rates Brownian motion model, and we inferred evolutionary
histories either including or eliminating the mean-centering
normalization step of our analysis pipeline. With the latter step in
place, our method correctly yielded little support for shifts in
evolutionary rates in the simulated data except in the case of
extremely tight correlation between genes, a regime unlikely to be
biologically relevant (Figure S7).  Additionally, to test the impact of our
assumption that the genes of a pathway were all subject to similar
evolutionary pressures, we simulated a heterogeneous pathway in which
expression of only a fraction of the gene members was subject to a
lineage-specific shift in evolutionary rate.  Inferring parameters
from these data revealed accurate detection of rate shifts even when a
large proportion of the genes in the pathway deviated from the rate
shift model (Figure S8).  Taken together, our results
make clear that the pathway-based phylogenetic approach is highly
powered to infer evolutionary histories of gene expression change,
particularly lineage-specific evolutionary rate shifts. As a
contrast to the poor performance of phylogenetic inference when
applied to one or a few genes, our findings underscore the utility of
the multi-gene paradigm in identifying candidate cases of
evolutionarily relevant expression divergence.

\subsection*{Phylogenetic inference of regulatory evolution from experimental measurements of \emph{Saccharomyces} expression}

We next set out to apply our method for evolutionary reconstruction of
regulatory change to experimental measurements of gene expression.
The total difference in gene expression between any two species is a
consequence of heritable differences that act in \emph{cis} on the DNA
strand of a gene whose expression is measured, and of variants that
act in \emph{trans}, through a soluble factor, to impact gene
expression of distal targets.  Effects of \emph{cis}-acting variation
can be surveyed on a genomic scale using our previously reported
strategy of mapping of RNA-seq reads to the individual alleles of a
given gene in a diploid inter-specific hybrid
\cite{bullard2010polygenic}, whereas the joint effects of \emph{cis}
and \emph{trans}-acting factors can be assessed with standard
transcriptional profiling approaches in cultures of purebred species.
To apply these experimental paradigms we chose a system of
\emph{Saccharomyces sensu stricto} yeasts.  We cultured two biological
replicates for each of a series of hybrids formed by the mating of
\emph{S. cerevisiae} to \emph{S. paradoxus}, \emph{S. mikatae}, and
\emph{S. bayanus} in turn, as well as homozygotes of each species.  We
measured total expression in the species homozygotes, and
allele-specific expression in the hybrids, of each gene by RNA-seq,
using established mapping and normalization procedures (see Methods).
In each set of expression data, we made use of \emph{S. cerevisiae} as
a reference: we normalized expression in the homozygote of a given
species, and expression of the allele of a given species in a diploid
hybrid, relative to the analogous measurement from
\emph{S. cerevisiae}.

To search for evidence of evolutionary constraint and lineage-specific
shifts in evolutionary rate in our yeast expression data, we
considered as pathways the pre-defined sets of genes of common
function from the Gene Ontology (GO) process categories.  For the
genes of each GO term, we used normalized expression measurements in
yeast species and, separately, measurements of \emph{cis}-regulatory
variation from interspecific hybrids, as input into our phylogenetic
analysis pipeline.  Thus, for each of the two classes of expression
measurements, for a given GO term we fit models of a lineage-specific
rate shift in regulatory evolution incorporating
inverse-gamma-distributed rates across genes; an analogous model with
no lineage-specific rate shift; and an OU model of universal
constraint.  The results revealed a range of inferred evolutionary
models and AIC support across GO terms (Figure 3, Tables 1 and 2, and
Tables S3 and S4), and this complete data set served as
the basis for manual inspection of biologically interesting features.

Among the inferences of pathway regulatory evolution from our method,
we observed many cases of evolutionary interest whose best-fitting
model had strong AIC support (Figure 3).  For each of 15 GO terms,
\emph{cis}-regulatory expression variation measurements yielded
inference of an evolutionary model with $>$80\% AIC weight (Figure 3a
and Table 1).  Many such GO terms represented candidate cases of
polygenic regulatory evolution, in which multiple independent
variants, at the unlinked genes that make up a pathway, have been
maintained in some yeast species in response to a lineage-specific
shift in selective pressure on expression of the pathway components.
For example, in replicative cell aging genes (GO term 0001302),
\emph{cis}-regulatory variation measured in interspecific hybrids
supported a model of polygenic, accelerated evolution in
\emph{S. paradoxus} (Figure 4a), with some pathway components
upregulated and some downregulated in the latter species relative to
other yeasts.  The total expression levels of cell aging genes in
species homozygotes were also consistent with rapid evolution in
\emph{S. paradoxus} (Figure 4a), arguing against a model of
compensation between \emph{cis}- and \emph{trans}-acting regulatory
variation, and highlighting this pathway as a particularly compelling
potential case of a lineage-specific change in selective pressure.

In other instances, expression measurements in species homozygotes
alone supported models of lineage-specific evolution, with each such
pathway representing a candidate case of accelerated or constrained
evolution at \emph{trans}-acting regulatory factors.  For a total of
41 GO terms, our method inferred models with $>$80\% AIC weight from
homozygote species expression data (Figure 3b and Table 2).  These
top-scoring pathways included a set of components of the transcription
machinery (GO term 0006351), whose expression levels in
\emph{S. bayanus} were less volatile than those of other yeasts and
thus supported a model of lineage-specific constraint (Figure 4b).
Additionally, expression of a number of pathways in species
homozygotes conformed to the OU model of universal constraint, such as
a set of genes annotated in transport (GO term 0006281), whose
expression varied less across all species than would be expected from
the genome tree (Figure 4c).  Taken together, our findings indicate
that evolutionary histories can be inferred with high confidence from
experimental measurements of pathway gene expression.  In our yeast
data, many pathways exhibit expression signatures consistent with
non-neutral regulatory evolution, in particular lineages and across
the phylogeny.

Another emergent trend was the prevalence, across many GO terms, of
models of distinct regulatory evolution in the lineage to
\emph{S. paradoxus} as the best fit to expression measurements in
species homozygotes (Figure 3b).  We noted no such recurrent model in
analyses of \emph{cis}-regulatory variation (Figure 3a), implicating
\emph{trans}-acting variants as the likely source of the regulatory
divergence in \emph{S. paradoxus}.  To validate these patterns, we
applied our phylogenetic inference method to expression measurements
from all genes in the genome analyzed as a single group, rather than
to each GO term in turn.  When we used expression data from species
homozygotes as input for this genome-scale analysis, our method
assigned complete AIC support to a model in which the rate of
evolution was $2.5$ times faster on the branch leading to
\emph{S. paradoxus} (AIC weight $= 1$), consistent with results from
individual GO terms (Figure 3b). An analogous inference calculation
using measurements of \emph{cis}-regulatory variation, for all genes
in the genome, yielded essentially complete support for an OU model of
universal constraint (AIC weight $= .99$).  We conclude that
constraint on the \emph{cis}-acting determinants of gene expression,
of roughly the same degree in all yeasts, is the general rule from
which changes in selective pressure on particular functions may drive
deviations in individual pathways.  However, for many genes,
expression in the \emph{S. paradoxus} homozygote is distinct from that
of other yeasts out of proportion to its sequence divergence,
suggestive of derived, \emph{trans}-acting regulatory variants with
pleotropic effects.

\section*{Discussion}

The effort to infer evolutionary histories of gene expression change
has been a central focus of modern comparative genomics.  Against a
backdrop of a few landmark successes \cite{ bedford2009optimization, 
brawand2011evolution}, progress in the field has been limited 
by the relatively weak power of phylogenetic methods when
applied, on a gene-by-gene basis, to measurements from small sets of
species.  In this work, we have met this challenge with a method to
infer evolutionary rates of any suite of independently measured
continuous characters that can be analyzed together across species.
We have derived the mathematical formalism for this model, and we have
illustrated the power and accuracy of our approach in simulations.  We
have generated yeast transcriptional profiles that complement
available data sets \cite{busby2011expression,goodman2013pervasive} by
measuring \emph{cis}-regulatory contributions to species expression
differences as well as the total variation between species.  With
these data, we have demonstrated that our phylogenetic inference
method yields robust, interpretable candidate cases of pathway
regulatory evolution from experimental measurements.

The defining feature of our phylogenetic inference method is that it
gains power by jointly leveraging expression measurements of a group
of genes, while avoiding a high-dimensional evolutionary model.
Rather than requiring an estimate of the evolutionary rate at each
gene, our strategy estimates the parameters of a distribution of
evolutionary rates across genes.  We thus apply the assumption of
\cite{chaix2008evolution} and model expression of the individual genes
of a pathway as independent draws from the same distribution,
mirroring the standard assumption of independence across sites in
phylogenetic analyses of DNA sequence \cite{felsenstein2004inferring}.
Any observation of lineage-specific \emph{cis}-acting regulatory
variation from our approach is of immediate evolutionary interest:  a
species-specific excess of variants at unlinked loci of common
function would be unlikely under neutrality, and would represent a
potential signature of positive selection if fixed across individuals
of the species.  In the study of \emph{trans}-acting regulatory variation,
\emph{a priori} a case of apparent accelerated evolution of a pathway
could be driven by a single mutation of large effect maintained by
drift in a species, as in any phenomenological analysis of trait
evolution \cite{lande1976natural, barton1989evolutionary}.  Our  results
indicate that for correlated gene groups, the latter issue can be largely resolved
by a simple transformation in which expression of each gene is normalized against
the mean of all genes in the pathway.  Additional corrections could be
required under more complex models of correlation among pathway genes,
potentially to be incorporated with matrix-regularization techniques
that highlight patterns of correlation in transcriptome data
\cite{dunn2013phylogenetic}.  Similarly, although the assumption of
independence across genes could upwardly bias the likelihoods of 
best-fit models in our inferences, model choice and parameter estimates will still be
correct on average even with the scheme implemented here \cite{varin2011overview}.

Our strategy also assumes that the genes of a pre-defined pathway are
subject to similar evolutionary pressures.  Simulation results
indicate that this assumption does not compromise the performance of
our method, as we observed robust inference to be the rule rather than
the exception even in a quite heterogeneous pathway, if a proportion
of the genes evolved under a rate shift model.  Although we have used
pathways defined by Gene Ontology in this study, our method can easily
be applied to gene modules defined on the basis of protein or genetic
interactions or coexpression.  Any such module is likely to contain
both activators and repressors, or other classes of gene function
whose expression may be quantitatively tuned in response to selection
by alleles with effects of opposite sign \cite{
  maughan2009transcriptome, howden2011evolution}.  The phylogenetic
approach we have developed here is well-suited to detect these
non-directional regulatory patterns, rather than relying on the
coherence of up- or down-regulation of pathway genes
\cite{bullard2010polygenic, fraser2010evidence,fraser2011systematic,
  fraser2012polygenic, martin2012evolution, fraser2013gene}.
Ultimately, a given case of strong signal in our pathway evolution
paradigm, when the best-fit model is one of lineage-specific
accelerated regulatory evolution, can be explained either as a product
of relaxed purifying selection or positive selection on pathway
output.  Our approach thus serves as a powerful strategy to identify
candidates for population-genetic \cite{martin2012evolution} and
empirical \cite{booth2010intercalation, fraser2012polygenic} tests of
the adaptive importance of pathway regulatory change.  We have
developed an R package, PIGShift (Polygenic Inverse Gamma rateShift),
to facilitate the usage of our method.  The pathway-level approach is
not contingent on the Gaussian models of regulatory evolution we have
used here, and future work will evaluate the advantages of compound
Poisson process \cite{khaitovich2005toward, chaix2008evolution} or
more general L\'{e}vy process \cite{landis2013phylogenetic} models of
gene expression.

The advent of RNA-seq has enabled expression surveys across non-model
species in many taxa.  Maximizing the biological value of these data
requires methods that evaluate expression variation in the context of
sequence divergence between species.  As rigorous phylogenetic
interpretation of expression data becomes possible, these measurements
will take their place beside genome sequences as a rich source of
hypotheses, in the search for the molecular basis of evolutionary
novelty.

\section*{Methods}
\subsection*{Basic model}

Our basic assumption, following \cite{chaix2008evolution}, is that the
average expression levels of genes in a pathway evolve as independent
replicates of the same Brownian motion or Ornstein-Uhlenbeck process.
However, instead of assuming that each gene in the pathway has the
same rate of evolution, we allow the different genes in a
pathway to draw their rate of evolution from a parametric
distribution.

As a point of departure, we begin by considering the likelihood of a
group of genes whose expression evolves independently, each with its
own rate of evolution.  Throughout, we use uppercase letters to
represent random variables and matrices and lowercase letters to
represent nonrandom variables.  Assume that we have measured
expression of the genes of a pathway in $n$ species, and that we have
a fixed, time-calibrated phylogeny from genome sequence data
describing the relationships between those species. We let
$\mathbf{X}_i = (X_{i,1}, X_{i,2}, \ldots, X_{i,n})$ be the
observations of the expression level of the $i$th gene of the pathway,
in each of $n$ species. Both the Brownian- motion and
Ornstein-Uhlenbeck (OU) models predict that the vector $\mathbf{X}_i$
is a draw from a multivariate normal distribution with
variance-covariance matrix $\sigma_i^2 \mathbf{V}$ (where $\sigma_i^2$
is a scalar---the rate of evolution---and the elements of $\mathbf{V}$
depend on whether evolution follows the Brownian or Ornstein-Uhlenbeck
model; see below). Hence, the likelihood of the data is

\begin{equation}
g(\mathbf{X}) = \prod_i\frac{1}{\sqrt{(2\pi\sigma_i^2)^n\det(\mathbf{V})}}e^{-\frac{1}{2\sigma_i^2}(\mathbf{x}_i-\mathbf{\mu}_i)'\mathbf{V}^{-1}(\mathbf{x}_i-\mathbf{\mu}_i)}
\end{equation}

where $\mathbf{\mu}_i$ is a vector representing the mean expression
value at the tips of the phylogenetic tree for gene $i$.  Note that
$\sigma_i^2V_{j,k} = \text{Cov}(X_{i,j}, X_{i,k})$ where $V_{i,j}$ is
the $i,j$th element of $\mathbf{V}$.

If we assume that there is no branch-specific directionality to
evolution, we can avoid the need to estimate $\mu$ in either the
Brownian motion model or the OU model by a renormalization of the
data. We first arbitrarily choose the gene expression measurements in
a single species (say species 1), and define the new random vector
$\mathbf{Z}_i = (Z_{i,2},Z_{i,3},\ldots,Z_{i,n})$ by

\[
Z_{i,j} = X_{i,j} - X_{i,1}.
\]

By our assumption that there is no branch-specific directionality,
$\mathbb{E}(X_{i,j}) = \mathbb{E}(X_{i,1})$ so $\mathbb{E}(Z_{i,j}) =
0$ for all $i$ and $j$. Because each $\mathbf{X}_i$ is multivariate
normally distributed with dimension $n$, each $\mathbf{Z}_i$ will also
be multivariate normally distributed with dimension $n-1$ and a
slightly different covariance structure. Letting $\mathbf{W}$ be the
covariance matrix corresponding to the $\mathbf{Z}_i$, elementary
calculations taking into account variances and covariances of sums of
random variables reveal that

\[
W_{i-1,j-1} = 
\begin{cases}
 V_{i,i} + V_{1,1} - 2V_{i,1} & \text{if } i = j \\
 V_{i,j} + V_{1,1} - V_{i,1} - V_{j,1} & \text{if }  i \neq j.
\end{cases}
\]

Next, we wish to incorporate into the Brownian motion and OU models a
scheme in which the rates of evolution of the genes of a pathway are
not specified independently but instead are drawn from an
inverse-gamma distribution. In this context, the genes in a pathway
share $\mathbf{W}$, the variance-covariance structure due to the
tree, but the rate of evolution $\sigma_i^2$ for each gene is an
independent draw from an inverse-gamma distribution.  The
inverse-gamma distribution has density

\begin{equation}
h(y) = \frac{\beta^\alpha}{\Gamma(\alpha)}y^{-(\alpha+1)}e^{-\frac{\beta}{y}},
\end{equation}

where $\Gamma(\cdot)$ is the gamma function and $\alpha$ and $\beta$
are shape and scale parameters. The moments of this distribution are

\[
\mathbb{E}(Y) = \frac{\beta}{\alpha-1}
\]
and
\[
\text{Var}(Y) = \frac{\beta^2}{(\alpha-1)^2(\alpha-2)},
\]

from which it follows that the inverse-gamma distribution has no mean
if $\alpha < 1$ and no variance if $\alpha < 2$. These properties
allow for the distribution of rates of gene expression evolution in a
pathway to be relatively broad; in addition, the inverse gamma
density has no mass at $0$, which prevents any gene in
a pathway from not evolving at all. Also, as $\alpha \rightarrow
\infty$ and $\beta \rightarrow \infty$ as $\frac{\beta}{\alpha-1} =
\mu$ stays fixed, the distribution converges to a point mass at
$\mu$. Thus, a model where there is one rate for every gene is nested
within the inverse-gamma distributed rates model.

Computation of the the likelihood of the data under this model is
simplified by the fact that the inverse-gamma distribution is the
conjugate prior to the variance of a normal distribution. Hence, we
see that the likelihood of the observed expression data $\mathbf{Z}$
is

\begin{eqnarray}
L(\mathbf{Z}) &=& \idotsint_0^\infty g(\mathbf{Z})h(\sigma_1^2)h(\sigma_2^2)\cdots h(\sigma_n^2)d(\sigma_1^2)d(\sigma_2^2 )\cdots d(\sigma_n^2)\nonumber \\
&=& \prod_i \int_0^\infty \frac{1}{\sqrt{(2\pi\sigma^2)^{n-1}\det(\mathbf{W})}}e^{-\frac{1}{2\sigma^2}\mathbf{z_i}'\mathbf{W}^{-1}\mathbf{z_i}}\frac{\beta^\alpha}{\Gamma(\alpha)}(\sigma^2)^{-(\alpha+1)}e^{-\frac{\beta}{\sigma^2}}d(\sigma^2) \nonumber \\
&=& \prod_i \frac{1}{\sqrt{(2\pi)^{n-1}\det(\mathbf{W})}}\frac{\beta^\alpha}{(\frac{1}{2}\mathbf{z}_i'\mathbf{W}^{-1}\mathbf{z}_i+\beta)^{\alpha+(n-1)/2}}\frac{\Gamma(\alpha + (n-1)/2)}{\Gamma(\alpha)}.
\label{like}
\end{eqnarray}

The second line follows recognizing that each integral is independent.
Thus, the likelihood of the observations of transcriptome-wide gene
expression across the pathway in $n$ taxa, normalized by the
expression level in taxon $1$, is given by \eqref{like}.

For the application to simulated and experimental data as described
below, given observations of gene expression of the species at the
tips of the tree, and a model that specifies the covariance matrix
$\mathbf{V}$ detailed in the next section, we optimized the log
likelihood function using the L-BFGS-B optimization routine in R
\cite{Zhu1997lbfgsb}.

\subsection*{Covariance matrix}

In the previous section, we left the unnormalized covariance matrix
$\mathbf{V}$ unspecified. Here we briefly recall the forms of
$\mathbf{V}$ under Brownian motion and the Ornstein-Uhlenbeck
process. Define the height of the evolutionary tree to be $T$ and and
the height of the node containing the common ancestor of taxa $i$ and
$j$ by $t_{ij}$. Then the covariance matrix for Brownian motion is

\[
V_{i,j} =
\begin{cases}
t_{ij} & \text{if } i \neq j \\
T & \text{if } i = j
\end{cases}
\]

and the covariance matrix for the Ornstein-Uhlenbeck process is

\[
V_{i,j} = 
\begin{cases}
\frac{1}{2\theta}e^{-2\theta(T-t_{ij})}(1-e^{2\theta t_{ij}}) & \text{if } i \neq j \\
\frac{1}{2\theta}(1-e^{2\theta T}) & \text{if } i = j
\end{cases}
\]

where $\theta$ quantifies the strength of stabilizing selection, with large
$\theta$ corresponding to stronger selection.

To model lineage-specific shifts in the evolutionary rate of gene
expression in the context of the Brownian motion model, we adopt a
framework similar to that of O'Meara \emph{et
  al.}\cite{o2006testing}. We assume that in a specified subtree of
the total phylogeny, the rate of evolution of every gene is multiplied
by a constant, compared to the rest of the tree. Under the Brownian
motion model, this is equivalent to multiplying the branch lengths in
that part of the tree by that same constant; hence, shifts in
evolutionary rate are incorporated by multiplying the branch lengths of affected branches by the
  value of the rate shift.

\subsection*{Comparing likelihoods among fitted models}

To evaluate the support for the distinct models we fit to expression
data for a given pathway, we require a strategy that will be broadly
applicable in cases where no \emph{a priori} expectation of the
correct model is available, such that nested hypothesis testing
schemes \cite{o2006testing} are not applicable.  Instead, given
likelihoods $L$ from fitting of each model in turn to expression data
from the genes of a pathway, we use the Akaike Information Criterion,
$2k - 2ln(L)$ \cite{akaike1974new}, to report the strength of the
support for each, where $k$ is the number of parameters in the model
($k = 2$ for the Brownian motion model in which the rate of evolution
is the same along all lineages in the phylogeny, and $k = 3$ for all
other models).

\subsection*{Simulations}

For all simulations, we used a phylogenetic tree adapted from
\cite{scannell2011awesome} by removing the branch leading to
\emph{Saccharomyces kudriavzevii} (see inset of Figure 1a and
Figures S1-S5).  To simulate under models in which each
gene in a pathway evolves independently, we generated expression data
for one gene at a time as follows.  We first drew the rate of
evolution from the appropriately parameterized inverse-gamma
distribution. Then, without loss of generality, we specified that the
expression level at the root of the phylogeny was equal to $0$, and we
simulated evolution along the branches of the yeast phylogeny
according to either a Brownian motion or an Ornstein-Uhlenbeck process
(with optimal expression level equal to $0$), using the terminal
expression level on a branch as the initial expression level of its
daughter branches.  To account for lineage-specific shifts in
evolutionary rate in a simulated pathway, we multiplied the rate of
evolution of each gene by the rate shift parameter for evolution along
the branches affected by the rate shift.  For each Brownian
motion-based rate shift model applicable to the tree, we simulated 100
replicate datasets for each of a range of gene group sizes, in each
case setting $\alpha = 3$, $\beta = 2$, and the rate shift parameter
as specified in Figure 1 and Figures S1-S5. For the
Ornstein-Uhlenbeck model, we simulated 100 replicate datasets for each
of a range of pathway sizes with $\alpha = 3$, $\beta = 2$, and
$\theta$ as specified in Figure 2.

To simulate under models in which expression of genes in a pathway was
correlated with coefficient $\rho$, we first drew $(\sigma^2_i, 1 \leq
i \leq n)$, the rate of evolution for each gene, from an inverse-gamma
distribution with $\alpha = 3$, $\beta = 2$. We then parameterized the
instantaneous variance-covariance matrix of the $n$-dimensional
Brownian motion by
 
\[
\Sigma_{i,j} = \begin{cases}
\sigma^2_i & \text{if } i = j \\
\rho\sigma_i\sigma_j & \text{if } i \neq j
\end{cases}
\]

so that the distribution of trait change along a lineage was
multivariate normal with mean 0 and variance covariance matrix
$\Sigma$.  Separate simulated expression data sets were generated with
$\rho$ varying from 0 (complete independence) to 1 (complete
dependence) using 100 replicate simulations for each value.

\subsection*{Yeast strains, growth conditions, and RNA-seq}

Strains used in this study are listed in Table S1.  For pairwise
comparisons of \emph{S. cerevisiae} and each of \emph{S. paradoxus},
\emph{S. mikatae}, and \emph{S. bayanus}, two biological replicates of
each diploid parent species and each interspecific hybrid were grown
at 25$^\circ$C in YPD medium \cite{ausubel2002short} to log phase
(between 0.65-0.75 OD at 600 nm). Total RNA was isolated by the hot
acid phenol method \cite{ausubel2002short} and treated with Turbo
DNA-free (Ambion) according to the manufacturer's
instructions. Libraries for a strand-specific RNA-seq protocol on the
Illumina sequencing platform, which delineates transcript boundaries
by sequencing poly-adenylated transcript ends, were generated as in
\cite{yoon2010noncanonical} with the following modifications: 1)
AmpureXP beads (Beckman) were used to clean up enzymatic reactions; 2)
the gel purification and size-selection step was eliminated; 3) the
oligo-dT primer used for cDNA synthesis was phosphorothioated at
position ten (TTTTTTTTTT*TTTTTTTTTTVN, V=A,C,G, N=A,C,G,T,
*=phosphorothioate linkage, Integrated DNA Technologies); and 4) 12
PCR cycles were performed. Libraries were sequenced using 36 bp
paired-end modules on an Illumina IIx Genome Analyzer (Elim
Biopharmaceuticals).

\subsection*{RNA-seq mapping and normalization}

Bioinformatic analyses were conducted in Python
 and R.  RNA-seq reads were stripped of their putative poly-A tails by removing
stretches of consecutive Ts flanking the sequenced fragment; reads
without at least two such Ts were discarded, as were reads with Ts at
both ends. To ensure that expression data from hybrid diploids and
purebred species could be compared, for each class of expression
measurement for a given pair of species we mapped reads to both
species genomes from http://www.saccharomycessensustricto.org
\cite{scannell2011awesome} using Bowtie \cite{langmead2009ultrafast} with
default settings and flags -m1 -X1000. These settings allowed us to
retain only those reads that were unambiguously assigned to one of the
two species in each pairwise comparison. A mapped read was inferred to
have originated from the plus strand of the genome if its poly-A tail
corresponded to a stretch of As at the 3$'$ end of the fragment, and a
read was assigned to the minus strand if its poly-A tail corresponded
to a stretch of Ts at the 5$'$ end of the fragment relative to the
reference genome. To filter out cases in which inferred poly-A tails
originated from stretches of As or Ts encoded endogenously in the
genome, we eliminated from analysis all reads whose stretch of As or
Ts contained more than 50\% matches to the reference genome. In order
to filter out cases of potential oligo-dT mispriming during cDNA
synthesis, we also eliminated from analysis all reads that contained
10 or more As in the 20 nucleotides upstream of their transcription
termination site. Read mapping statistics can be found in Table S2.

We controlled for read abundance biases due to differing GC content as
follows. For each lane of sequencing, we grouped sets of overlapping
reads and normalized abundance according to GC content of the
overlapping region using full-quantile normalization as implemented in
the package EDASeq \cite{risso2011gc}. Normalized abundance was
divided by raw abundance to generate a weight that was assigned to
every read in the group. These weights were used in place of raw read
counts in all downstream analyses.  All expression data are available
through the Gene Expression Omnibus under identification number
GSE38875.

\subsection*{Transcript annotation}

Coordinates of orthologous open reading frames (ORFs) in each genome
were taken from \\ 
\texttt{http://www.saccharomycessensustricto.org}. These ORF
boundaries in \emph{S. cerevisiae} differed, in some cases, from ORF
definitions in the \emph{Saccharomyces} Genome Database \cite[SGD,
  using the definitions from December 22, 2007]{cherry1998sgd}; genes
for which the two sets of definitions did not overlap were
discarded. For cases where the definitions overlapped but differed by
more than ten base pairs at either end, we used the boundaries defined
by SGD and adjusted ortholog boundaries in other species accordingly
after performing local multiple alignment \cite{edgar2004muscle} of
the orthologous regions and flanking sequences as defined by
\cite{scannell2011awesome}.

For most genomic loci, each sense transcript feature was defined as
the region from 50 bp upstream to 500 bp downstream of its respective
ORF. If sequence within this window for a given target ORF overlapped
with the boundaries of an adjacent gene or known non-coding RNA on the
same strand, the sense feature boundaries of the target were trimmed
to eliminate the overlap. For tandem gene pairs, the 3$'$ boundary of
the upstream gene sense feature was set to 500 bp past the coding
stop or the coding start of the downstream gene sense feature,
whichever was closer; the 5$'$ boundary of the downstream gene sense
feature was set to 50 bp upstream of its coding start or the 3$'$ end of
the upstream gene sense feature, whichever was closer.

We tabulated the GC-normalized expression counts (see above) that
mapped to each transcript feature for each RNA-seq sample. Given the
full set of such counts across all features and all samples, we then
applied the upper-quartile between-lane normalization method
implemented in EDASeq \cite{risso2011gc}. The normalized counts from
this latter step for a given species were averaged across all
biological replicates to yield a final expression level for the
feature, which we then $\log_2$ transformed and used in all analysis in this work.

\subsection*{Yeast pathways}

We downloaded the list of genes associated with each Gene Ontology
process term from the \emph{Saccharomyces} Genome Database and
filtered for terms containing at least 10 genes.  The resulting set
comprised 333 terms.

\subsection*{Visualizing distributions of interspecific expression variation}

For visual inspection of expression differences between species in
Figure 4, we normalized experimentally measured data by branch lengths
ascertained from genome sequence as follows.  If expression evolution
follows the same Gaussian-based model on all lineages of the yeast
phylogeny, when the expression level of gene $j$ in taxon $i$ is
compared to that in taxon $1$ used as a reference, the marginal
distribution $Z_{i,j}$ (the difference in expression between taxon $i$
and taxon $1$ at gene $j$) is distributed according to a univariate
analog of equation \eqref{like}. In this case, dividing $Z_{i,j}$ by
the absolute branch length according to DNA sequence between taxon $i$
and taxon $1$ eliminates the dependence of the distribution on the
total divergence time between taxa, and the density of this normalized
quantity will be the same for all species comparisons. In the case of
lineage-specific shifts in evolutionary rate or universal selective
constraint, one or more taxa will exhibit distinct densities of the
normalized expression divergence measure.  Thus, we generated each
distribution in Figure 4 by tabulating the log fold-change in
expression between the indicated species and \emph{S. cerevisiae}, and
then dividing this quantity by the divergence time between the
indicated species and \emph{S. cerevisiae} according to the genome
tree.  After this normalization, if a pathway has been subject to
accelerated regulatory evolution in one lineage, the distribution of
expression log fold-changes corresponding to the species at the tip of
that lineage will be wider than expected based on the length of the
branch from DNA sequence, and hence it will stand out against the
other distributions when plotted as in Figure 4; likewise, constraint
on expression evolution of a pathway in a particular species will
manifest as a narrower distribution for that species.  In the case of
a pathway subject to the same degree of regulatory constraint on all
branches of the yeast phylogeny, branch lengths ascertained from
genome sequence will be large relative to the modest expression
divergence, with the most dramatic disparity manifesting when
divergent species are compared, yielding the narrowest distribution of
normalized expression levels.  When visualized as in
Figure 4, the width of the distribution of log fold-changes across
genes of the pathway in a given species will thus be inversely
proportional to the species distance from \emph{S. cerevisiae}, with
the narrowest distribution for \emph{S. bayanus} and the widest for
\emph{S. paradoxus}.

\section*{Acknowledgments}
The authors thank Davide Risso and Oh Kyu Yoon for generously
providing software before publication; Daniela Delnieri, Chris Todd
Hittinger and Oliver Zill for providing \emph{Saccharomyces} strains;
and John Huelsenbeck, Mason Liang, Nicholas Matzke, Rasmus Nielsen,
Benjamin Peter, Jeremy Roop, and Montgomery Slatkin for helpful
discussions.

\bibliography{references}

\newpage
\section*{Figure Legends}

\begin{figure}[!htp]
\includegraphics{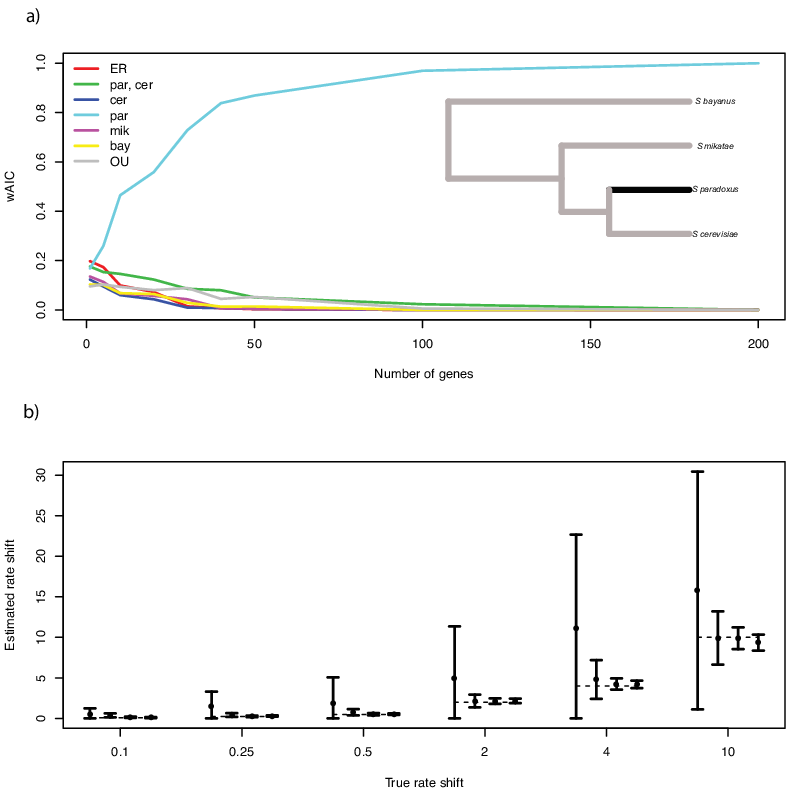}
\caption{{\bf Phylogenetic inference of the evolutionary history of
    yeast pathway regulation from data simulated under a model of a
    lineage-specific, accelerated evolutionary rate.}  Each panel
  reports results of the inference of evolutionary history from
  expression of the genes of a pathway in yeast species, simulated
  under a model of a shift in evolutionary rate on the branch leading
  to \emph{S. paradoxus} (dark line in inset phylogeny in (a)). (a),
  Each trace reports the strength of support for one evolutionary
  model in inferences from simulated expression in pathways of varying
  size.  The $x$ axis reports the number of genes in the pathway and
  the $y$ axis reports the Akaike weight of the indicated model.  Data
  were simulated under a Brownian motion model in which the rate of
  regulatory evolution for each gene was drawn from an inverse-gamma
  distribution with $\alpha = 3$, $\beta = 2$ and, for the branch
  leading to \emph{S. paradoxus}, increased by a factor of $5$. In the
  legend, ER denotes an equal-rates Brownian motion model in which
  rates of evolution were the same on each branch of the phylogeny; OU
  denotes an Ornstein-Uhlenbeck model of evolution; and species name
  abbreviations denote Brownian motion models of accelerated
  evolutionary rate on the subtrees leading to the respective
  taxa. (b), Each set of symbols reports results from expression data
  simulated under a Brownian motion model in which the rate of
  regulatory evolution for each gene was drawn from an inverse-gamma
  distribution with $\alpha = 3$, $\beta = 2$ and, for the branch
  leading to \emph{S. paradoxus}, increased by the factor indicated on
  the $x$ axis.  In a given set of symbols, filled circles report the
  mean, and vertical bars report the standard deviation of the
  sampling distribution, of the inferred rate shift parameter in
  simulations of pathways containing, from left to right, 2, 10, 50,
  and 100 genes.  Results from simulations of expression under models
  of evolutionary rate shifts on other branches of the yeast
  phylogeny, and simulations of expression in the absence of a
  lineage-specific evolutionary rate shift, are reported in
  Supplmentary Figures 1-5.}

\end{figure}

\newpage
\begin{figure}[!htp]
\includegraphics{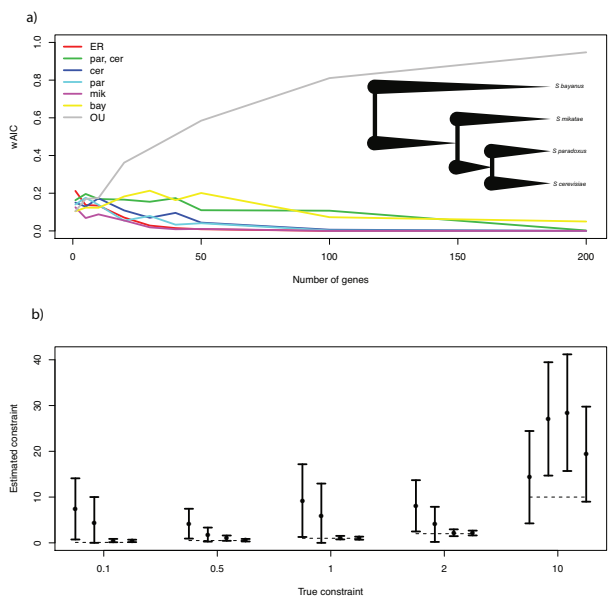}
\caption{{\bf Phylogenetic inference of the evolutionary history of yeast
pathway regulation from data simulated under an Ornstein-Uhlenbeck
(OU) model. } (a), Data are as in Figure 1a except that expression
measurements were simulated under an OU model in which the
phylogeny-wide rate of regulatory evolution for each gene was drawn
from an inverse-gamma distribution with
$\alpha = 3$, $\beta = 2$ and the phylogeny-wide constraint parameter had a value of 10. (b),
Data are as in Figure 1b except that expression measurements were
simulated under an OU model in which the phylogeny-wide rate of
regulatory evolution for each gene was drawn from an inverse-gamma
distribution with $\alpha = 3$, $\beta = 2$ and the phylogeny-wide
constraint parameter had the value indicated on the $x$ axis.
}

\end{figure}

\newpage
\begin{figure}[!htp]

\includegraphics{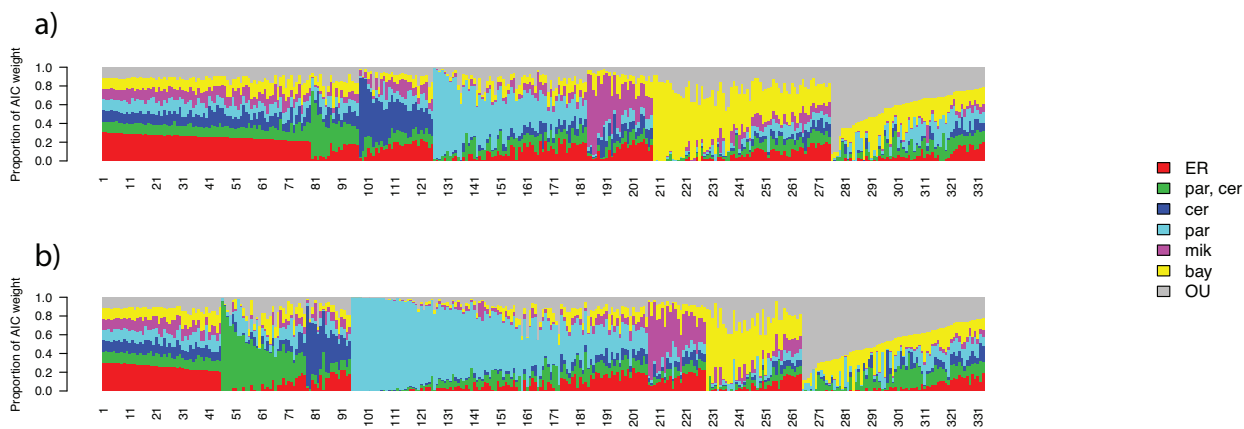}
    \caption{{\bf Inference of regulatory evolution in yeast
        pathways from experimental expression measurements.}  Each panel reports results of phylogenetic
      inference of evolutionary histories of gene expression change
      from one set of experimental transcriptional profiling data.  In
      a given panel, each vertical bar reports results of
      maximum-likelihood fits of Brownian-motion and
      Ornstein-Uhlenbeck models to expression of the genes of one Gene
      Ontology process term; the total proportion of a bar
      corresponding to a particular color indicates the Akaike weight
      of the corresponding model (legend at right, with labels as in
      Figure 1). Bars are sorted by the model with maximum Akaike
      weight. (a), Inference of \emph{cis}-regulatory variation from
      interspecies hybrids; numerical indices correspond to rows in
      Table S3.  (b), Inference from measurements of total
      expression in species homozygotes;
      numerical indices correspond to rows in Table S4.}
\end{figure}

\newpage
\begin{figure}[!htp]
\includegraphics{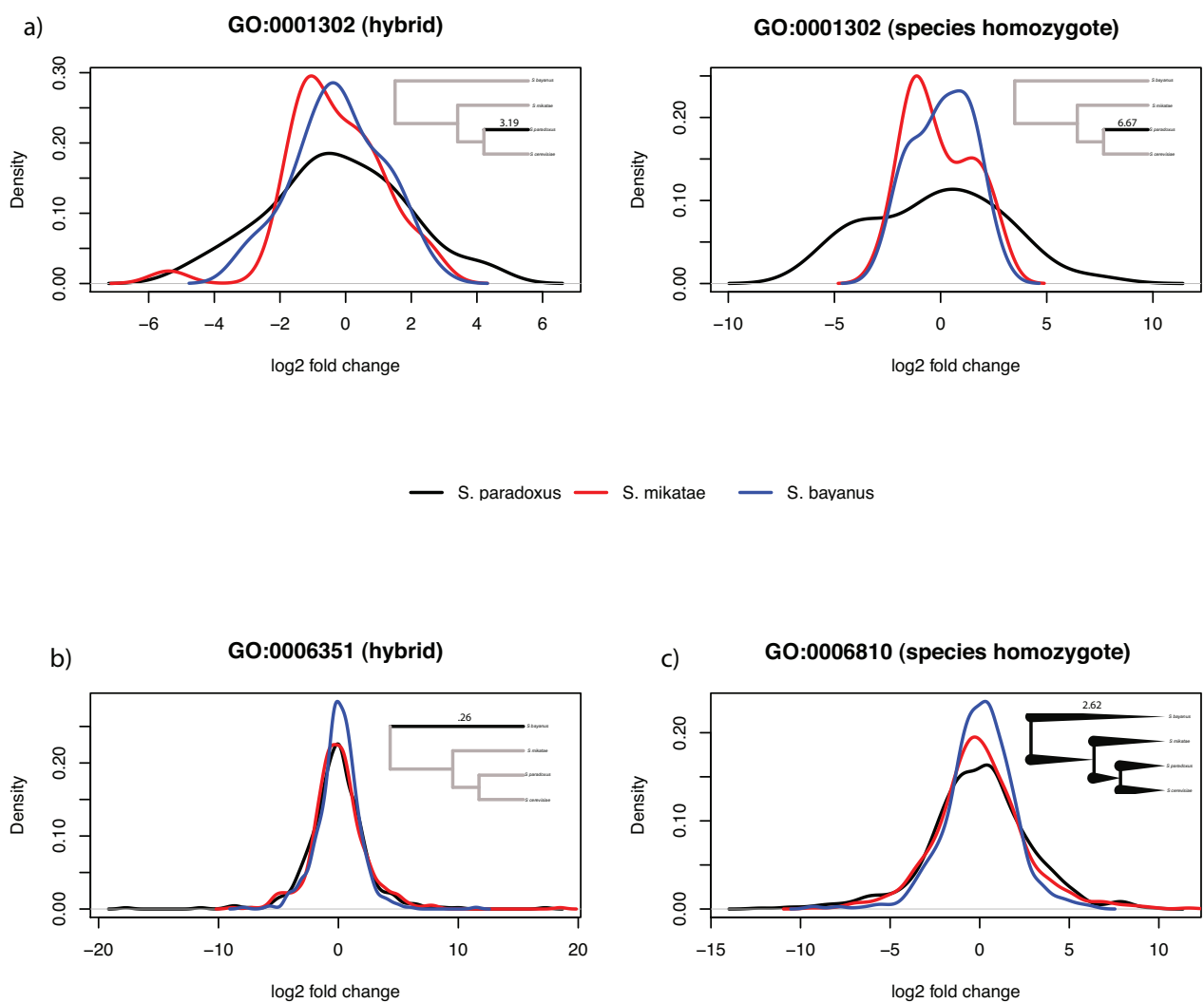}
\caption{{\bf Lineage-specific regulatory evolution and constraint in
    yeast pathways, inferred from experimental expression measurements.}  Each panel shows kernel density estimates of the
  distributions of experimental gene expression measurements among the
  genes of one yeast Gene Ontology process term, whose evolutionary
  history was inferred with strong support.  In a given panel, each
  trace reports the expression levels of the genes of the indicated
  pathway, from the allele of the indicated yeast species in a hybrid
  or in the purebred homozygote of a species, normalized with respect
  to the analogous measurement in \emph{S. cerevisiae} and with
  respect to branch length.  Inset cartoons represent the model
  inferred with AIC weight $>$80\% for the indicated pathway (see
  Tables 1 and 2).  (a) Allele-specific expression from measurements
  in diploid hybrids (left) and total expression measurements in
  species homozygotes (right) for the 38 genes of GO:0001302,
  replicative cell aging, supporting a model of accelerated evolution
  in \emph{S. paradoxus}; in the inset, the number above the bolded
  branch reports the inferred shift in the rate of regulatory
  evolution along that lineage.  (b) Allele-specific expression from
  measurements in diploid hybrids for the 462 genes of GO:0006351,
  transport, supporting a model of constraint in \emph{S. bayanus}; in
  the inset, the number above the bolded branch reports the inferred
  shift in the rate of regulatory evolution along that lineage.  (c)
  Total expression measured in species homozygotes for the 175 genes
  of GO:0006281, transcription, supporting an Ornstein-Uhlenbeck model
  of universal constraint; in the inset, the number above the tree
  reports the inferred value of the constraint parameter.  Note that
  in (c), the width of the distribution of expression differences
  between a given species and \emph{S. cerevisiae} correlates
  inversely with the sequence divergence of that species, as expected
  if selective constraint on expression renders the estimate of
  evolutionary distance from genome sequence an increasing
  over-estimate of expression change.}

\end{figure}

\bigskip
\newpage
\section*{Tables}

\begin{table}[!htp]
  \caption{
  \bf{Top-scoring fitted models of \emph{cis}-regulatory
      evolution in yeast pathways from experimental expression measurements.}
      }
    \begin{tabular}{lcccr}
    GO term & $N$ & Model & wAIC & Constraint or shift parameter \\ \hline
    34599 & 57 & Ornstein-Uhlenbeck & 0.899405768 & 49.97745883 \\
    6355 & 433 & \emph{S. bayanus} shift & 0.837382338 & 0.230918849 \\
    6351 & 462 & \emph{S. bayanus} shift & 0.849912647 & 0.258701476 \\
    1302 & 38 & \emph{S. paradoxus} shift & 0.859866949 & 3.197059161 \\
    6897 & 73 & \emph{S. paradoxus} shift & 0.965743399 & 4.292287639 \\
    6338 & 45 & \emph{S. cerevisiae} shift & 0.840339574 & 0.037806902 \\
    42254 & 136 & Ornstein-Uhlenbeck & 0.924785133 & 3.733770466 \\
    6364 & 177 & Ornstein-Uhlenbeck & 0.902358815 & 3.079387696 \\
    44255 & 13 & \emph{S. paradoxus} shift & 0.945799302 & 11.43989834 \\
    54 & 11 & \emph{S. paradoxus} shift & 0.91523272 & 9.314688245 \\
    16310 & 188 & \emph{S. bayanus} shift & 0.902247359 & 0.188381056 \\
    8152 & 243 & \emph{S. bayanus} shift & 0.844716856 & 0.043114988 \\
    6629 & 136 & \emph{S. bayanus} shift & 0.91650274 & 0.005082617 \\
    122 & 71 & \emph{S. bayanus} shift & 0.819216472 & 0.040060263 \\
    30437 & 45 & \emph{S. paradoxus} shift & 0.931136455 & 4.060128813 \\
    \end{tabular}

    \begin{flushleft}
    Each row reports the results of phylogenetic inference of the
    evolutionary history of gene regulation for one yeast Gene
    Ontology process term, from experimental measurements of
    \emph{cis}-regulatory variation in interspecific yeast hybrids.
    $N$, number of genes in the indicated GO term for which expression
    measurements were available in all species. Model, best-fit model
    from among the five possible Brownian motion models of
    evolutionary rate shift in lineages of the \emph{Saccharomyces}
    phylogeny (see Figure 1a), the Ornstein-Uhlenbeck (OU) model of
    universal constraint, and the equal-rates model involving no
    lineage-specific differences in evolutionary rate.  wAIC, Akaike
    Information Criterion weight of the indicated model.  Constraint
    or shift parameter, fitted value of the strength of purifying
    selection or the shift in the rate of regulatory evolution on the
    indicated lineage, when the best-fit model was the OU model of
    constraint or a Brownian motion lineage-specific evolutionary rate
    model, respectively.
      \end{flushleft}

\end{table}

\newpage
\begin{table}[!htp]
  \caption{
  \bf{Top-scoring fitted models of species regulatory evolution in
  yeast pathways from experimental expression measurements.}
  }
    \begin{tabular}{l c c c r}
    GO term & $N$ & Model & wAIC & Constraint or shift parameter \\ \hline
    6397 & 151 & \emph{S. paradoxus} shift & 0.965171603 & 3.028130303 \\
    8033 & 69 &\emph{S. paradoxus} shift & 0.969683391 & 3.714749932 \\
    71038 & 15 &\emph{S. paradoxus} shift & 0.89725301 & 6.751073973 \\
    480 & 29 &\emph{S. paradoxus} shift & 0.928296518 & 4.460579672 \\
    42274 & 25 &\emph{S. paradoxus} shift & 0.958076119 & 8.083546161 \\
    472 & 31 &\emph{S. paradoxus} shift & 0.953733629 & 4.686648741 \\
    15031 & 362 &\emph{S. bayanus} shift & 0.872939854 & 0.183834463 \\
    1302 & 38 &\emph{S. paradoxus} shift & 0.999927135 & 6.671016575 \\
    6006 & 22 &\emph{S. paradoxus} shift & 0.816341854 & 4.6555377 \\
    6260 & 72 &\emph{S. paradoxus} shift & 0.831407464 & 3.043207869 \\
    30163 & 15 &\emph{S. paradoxus} shift & 0.82364567 & 7.009201233 \\
    6897 & 73 &\emph{S. paradoxus} shift & 0.970677101 & 4.408614609 \\
    6412 & 228 &\emph{S. paradoxus} shift & 0.981277345 & 2.770778823 \\
    7121 & 16 &\emph{S. paradoxus} shift & 0.998579562 & 16.81960721 \\
    6914 & 49 &Ornstein-Uhlenbeck & 0.810293525 & 41.38598192 \\
    30488 & 18 &\emph{S. paradoxus} shift & 0.893282646 & 7.945094861 \\
    42254 & 163 &\emph{S. paradoxus} shift & 0.99999983 & 6.856141937 \\
    6200 & 34 &\emph{S. paradoxus} shift & 0.81144199 & 5.590943868 \\
    6468 & 120 &\emph{S. paradoxus} shift & 0.990399439 & 2.655209273 \\
    16567 & 71 &\emph{S. paradoxus} shift & 0.959694914 & 3.313920599 \\
    6364 & 177 &\emph{S. paradoxus} shift & 0.999995709 & 5.841035759 \\
    6754 & 18 &\emph{S. paradoxus} shift & 0.816303046 & 4.668929462 \\
    422 & 27 &Ornstein-Uhlenbeck & 0.877576591 & 57.08946364 \\
    463 & 20 &\emph{S. paradoxus} and \emph{S. cerevisiae} shift & 0.958484282 & 10.39289039 \\
    6414 & 23 &\emph{S. paradoxus} and \emph{S. cerevisiae} shift & 0.906687775 & 8.121469425 \\
    19236 & 29 &\emph{S. paradoxus} shift & 0.989881765 & 6.821984459 \\
    31505 & 72 &\emph{S. paradoxus} shift & 0.955855579 & 3.032267535 \\
    32259 & 65 &\emph{S. paradoxus} shift & 0.998665437 & 4.546902844 \\
    6506 & 29 &\emph{S. paradoxus} shift & 0.982054204 & 5.468542886 \\
    16310 & 188 &\emph{S. paradoxus} shift & 0.99652632 & 2.487101867 \\
    447 & 39 &\emph{S. paradoxus} shift & 0.994506418 & 5.252074336 \\
    6281 & 175 &Ornstein-Uhlenbeck & 0.882367142 & 3.410968446 \\
    71042 & 13 &\emph{S. paradoxus} shift & 0.804318406 & 6.030946867 \\
    6378 & 18 &\emph{S. cerevisiae} shift & 0.845112064 & 1.00E-04 \\
    7165 & 63 &\emph{S. paradoxus} shift & 0.811091269 & 4.465389345 \\
    6810 & 681 &Ornstein-Uhlenbeck & 0.859937275 & 2.618523967 \\
    6812 & 28 &\emph{S. paradoxus} shift & 0.898839416 & 4.312524185 \\
    8150 & 723 &\emph{S. paradoxus} shift & 0.999962114 & 2.871955612 \\
    6417 & 45 &\emph{S. paradoxus} shift & 0.925463092 & 5.339113187 \\
    6407 & 18 &\emph{S. paradoxus} shift & 0.988260506 & 8.792447836 \\
    462 & 55 &\emph{S. paradoxus} shift & 0.817627126 & 7.291083934 \\
\end{tabular}

\begin{flushleft}Data are as in Table 1 except that inferences were
  made from experimental measurements of expression in purebred yeast
  homozygotes.
  \end{flushleft}

\end{table}


\newpage
\section*{Supplementary Figure Legends}

\newpage
\begin{figure}[!htp]

\includegraphics[width=.8\textwidth]{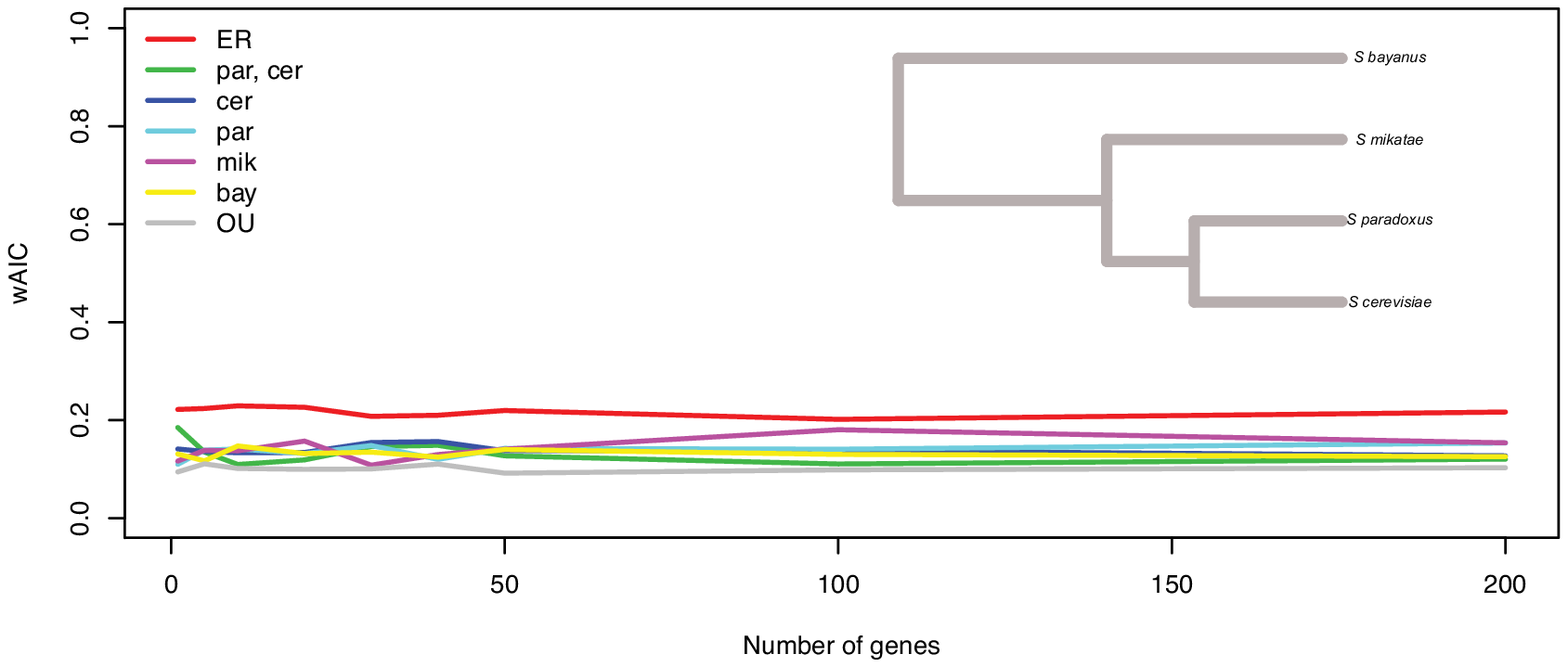}
\caption*{{\bf Figure S1. Phylogenetic inference of the
    evolutionary history of yeast pathway regulation under a Brownian
    motion model with equal rates on each branch of the tree.} Data
  are as in Figure 1a of the main text except that expression data
  were simulated under a model in which no yeast lineage was subject
  to a change in evolutionary rate.}

\end{figure}
\newpage

\begin{figure}[!htp]

\includegraphics[width=.8\textwidth]{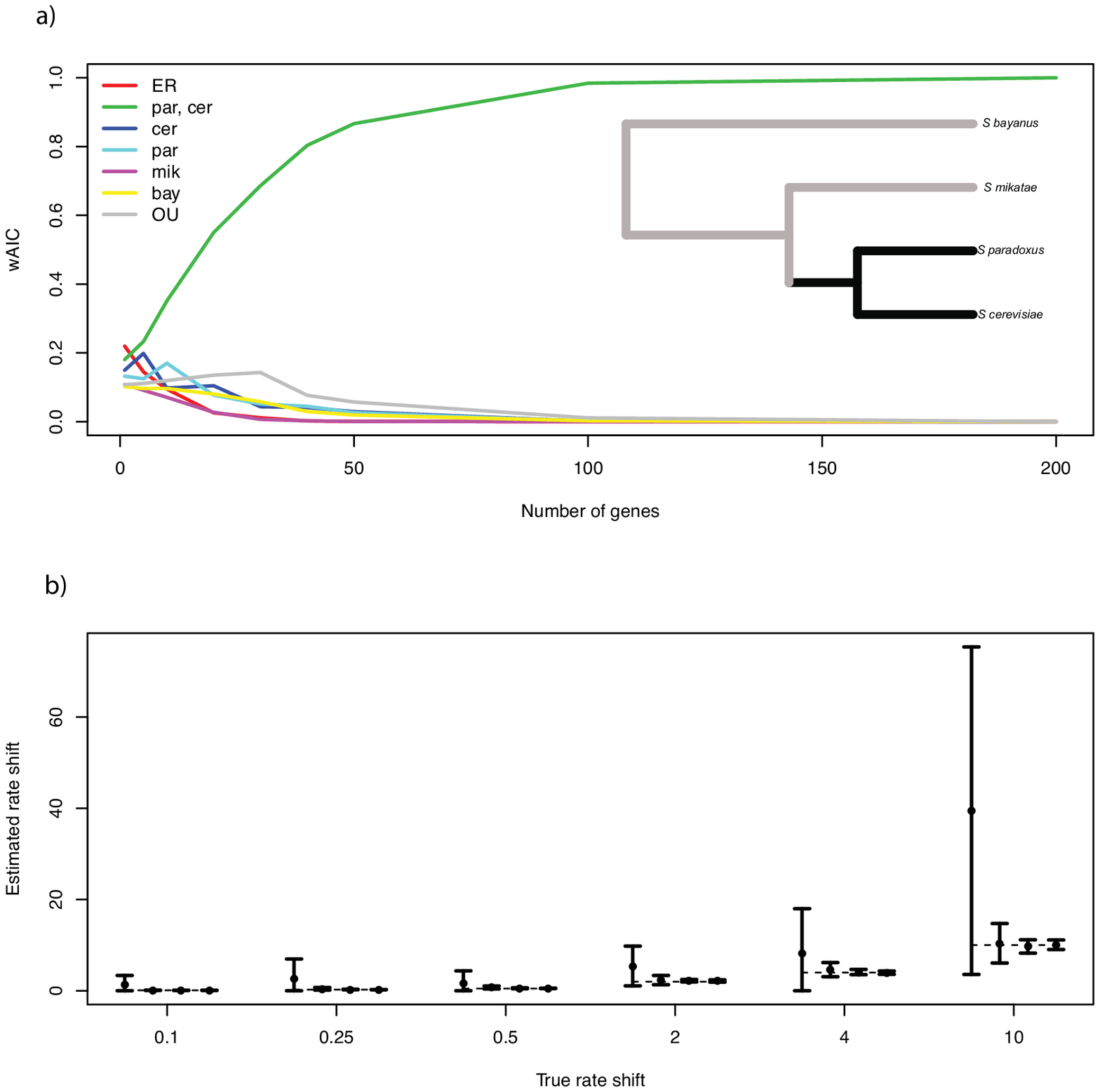}
\caption*{{\bf Figure S2. Phylogenetic inference of the
    evolutionary history of yeast pathway regulation under a model
    with a rate shift on the subtree leading to \emph{S. paradoxus}
    and \emph{S. cerevisiae}.} (a), Data are as in Figure 1a of the main text, except that expression
measurements were simulated under a Brownian motion model in which the
rate of regulatory evolution for each gene was drawn from an
inverse-gamma distribution with $\alpha$ = 3, $\beta$ = 2 and, for
the subtree leading to \emph{S. paradoxus} and \emph{S. cerevisiae},
increased by a factor of 5. (b), Data are as in Figure 1b of the main
text, except that expression measurements were simulated under a
Brownian motion model in which the rate of regulatory evolution for
each gene was drawn from an inverse-gamma distribution with 
$\alpha$ = 3, $\beta$ = 2 and, for the subtree leading to \emph{S. paradoxus}
and \emph{S. cerevisiae}, increased by the factor indicated on the $x$
axis.}

\end{figure}
\newpage

\begin{figure}[!htp]

\includegraphics[width=.8\textwidth]{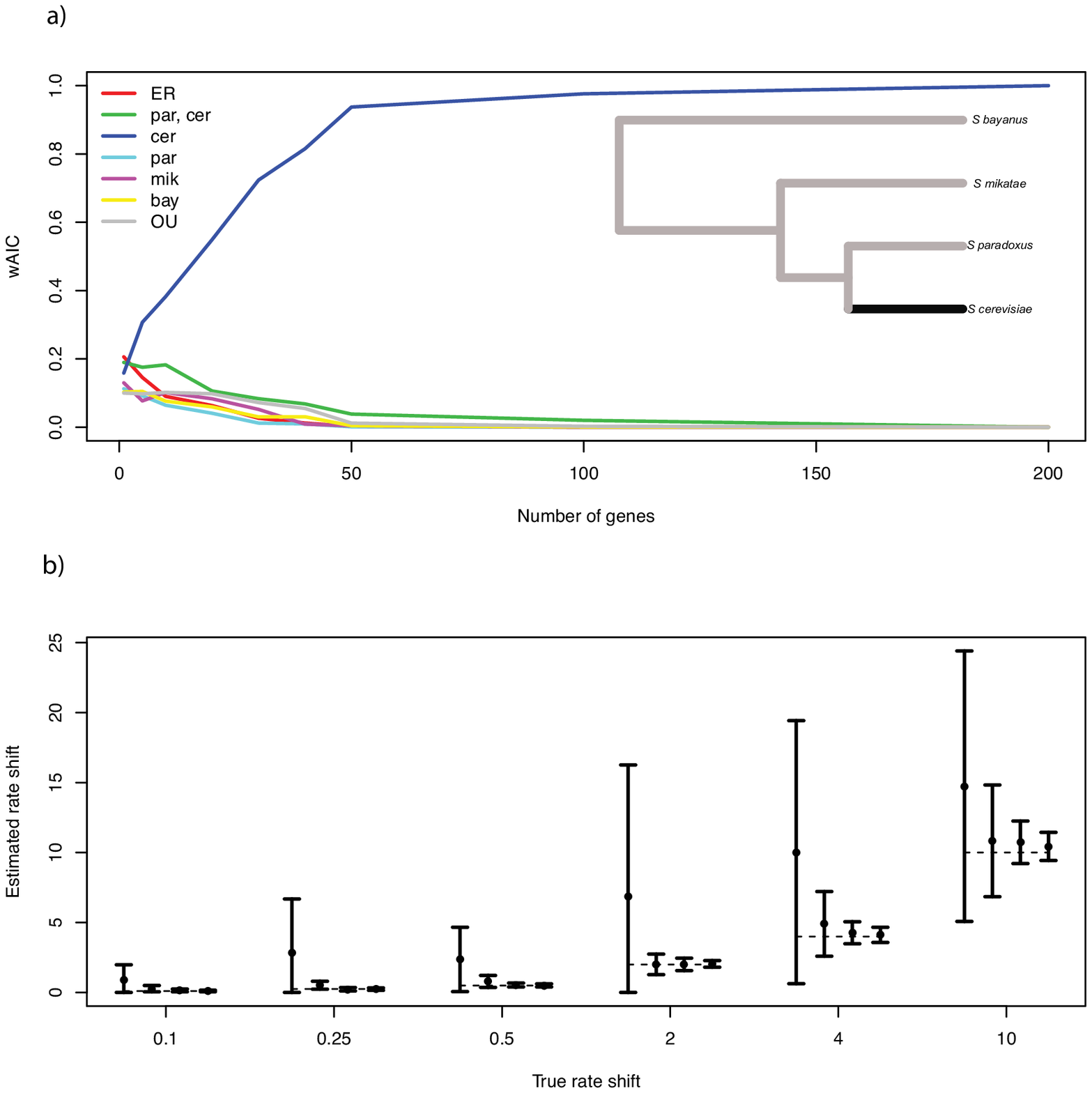}
\caption*{{\bf Figure S3. Phylogenetic inference of the
    evolutionary history of yeast pathway regulation under a model
    with a rate shift on the branch leading to \emph{S. cerevisiae}.}
  (a), Data are as in Figure
1a of the main text, except that expression measurements were
simulated under a Brownian motion model in which the rate of
regulatory evolution for each gene was drawn from an inverse-gamma
distribution with 
$\alpha$ = 3, $\beta$ = 2 and, for the branch
leading to \emph{S. cerevisiae}, increased by a factor of 5. (b), Data
are as in Figure 1b of the main text, except that expression
measurements were simulated under a Brownian motion model in which the
rate of regulatory evolution for each gene was drawn from an
inverse-gamma distribution with $\alpha$ = 3, $\beta$ = 2 and, for
the branch leading to \emph{S. cerevisiae}, increased by the factor
indicated on the $x$ axis.}

\end{figure}
\newpage

\begin{figure}[!htp]

\includegraphics[width=.8\textwidth]{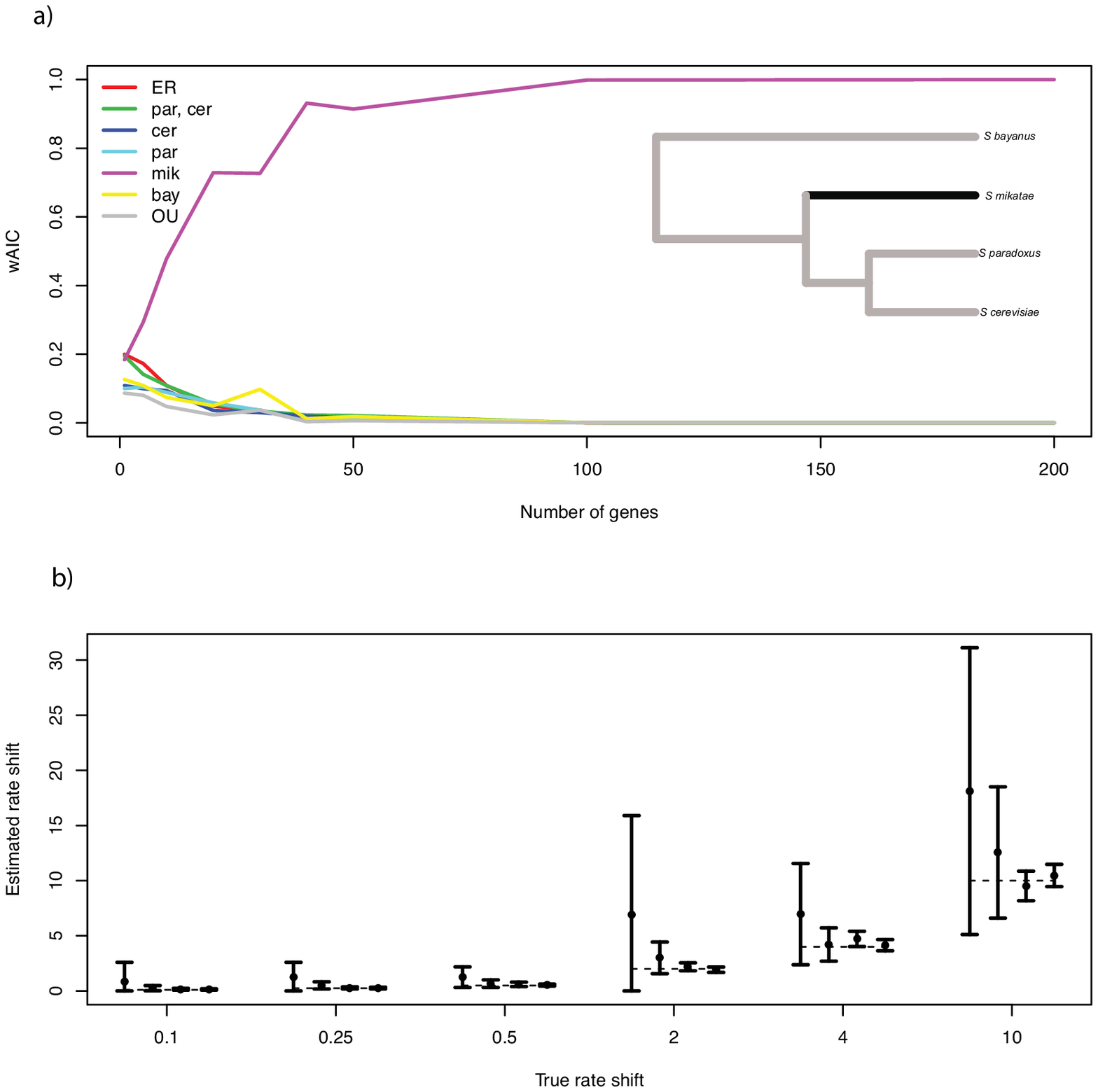}
\caption*{{\bf Figure S4. Phylogenetic inference of the
    evolutionary history of yeast pathway regulation under a model
    with a rate shift on the branch leading to \emph{S. mikatae}.}
  (a), Data are as in Figure 1a
of the main text, except that expression measurements were simulated
under a Brownian motion model in which the rate of regulatory
evolution for each gene was drawn from an inverse-gamma distribution
with $\alpha$ = 3, $\beta$ = 2 and, for the branch leading to
\emph{S. mikatae}, increased by a factor of 5. (b), Data are as in
Figure 1b of the main text, except that expression measurements were
simulated under a Brownian motion model in which the rate of
regulatory evolution for each gene was drawn from an inverse-gamma
distribution with $\alpha$ = 3, $\beta$ = 2 and, for the branch
leading to \emph{S. mikatae}, increased by the factor indicated on the
$x$ axis.}

\end{figure}
\newpage

\begin{figure}[!htp]

\includegraphics[width=.8\textwidth]{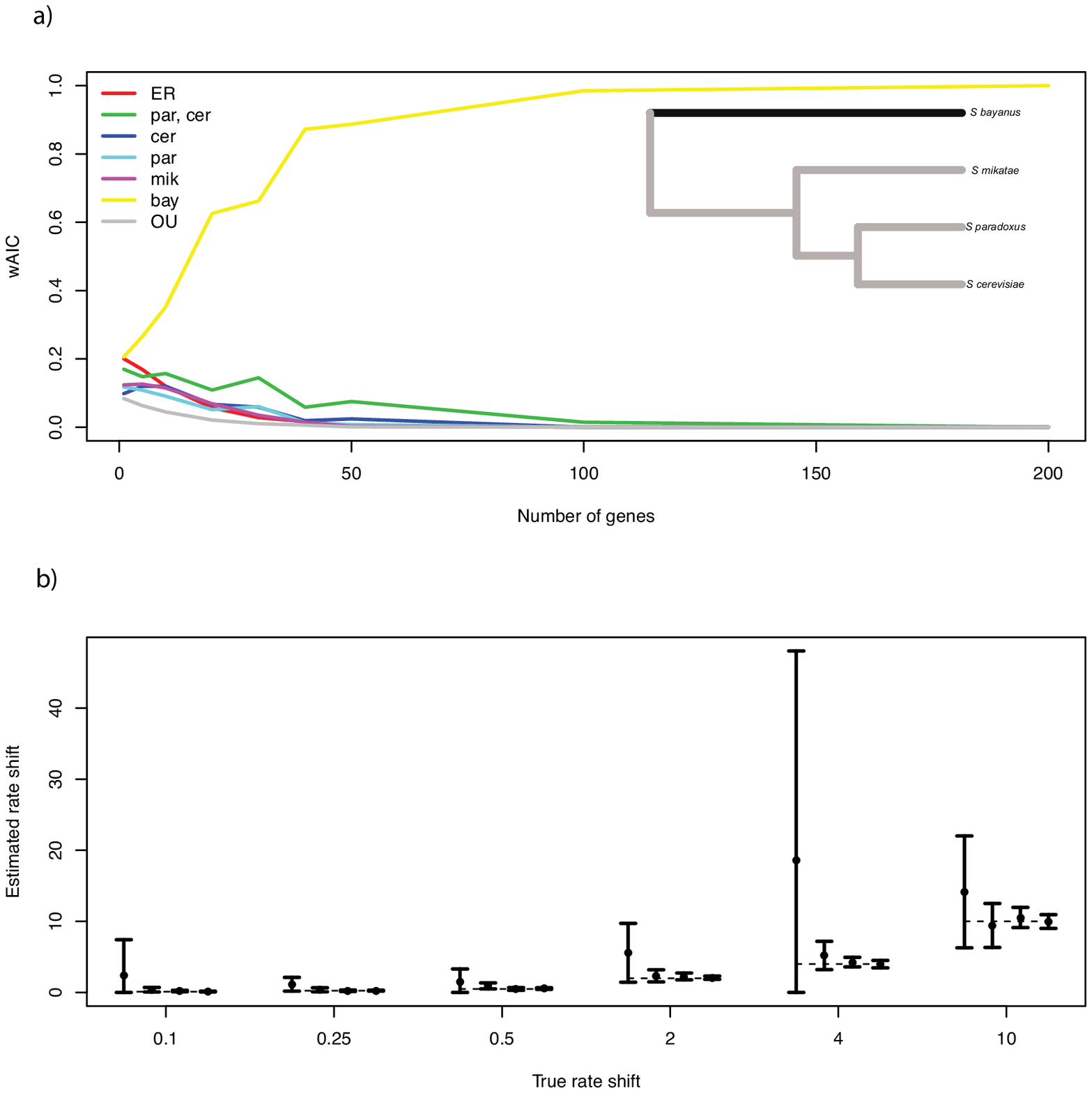}
\caption*{{\bf Figure S5. Phylogenetic inference of the
    evolutionary history of yeast pathway regulation under a model
    with a rate shift on the branch leading to \emph{S. bayanus}.}
  (a), Data are as in Figure 1a
of the main text, except that expression measurements were simulated
under a Brownian motion model in which the rate of regulatory
evolution for each gene was drawn from an inverse-gamma distribution
with $\alpha$ = 3, $\beta$ = 2 and, for the branch leading to
\emph{S. bayanus}, increased by a factor of 5. (b), Data are as in
Figure 1b of the main text, except that expression measurements were
simulated under a Brownian motion model in which the rate of
regulatory evolution for each gene was drawn from an inverse-gamma
distribution with $\alpha$ = 3, $\beta$ = 2 and, for the branch
leading to \emph{S. bayanus}, increased by the factor indicated on the
$x$ axis.}

\end{figure}
\newpage

\includegraphics[width=.8\textwidth]{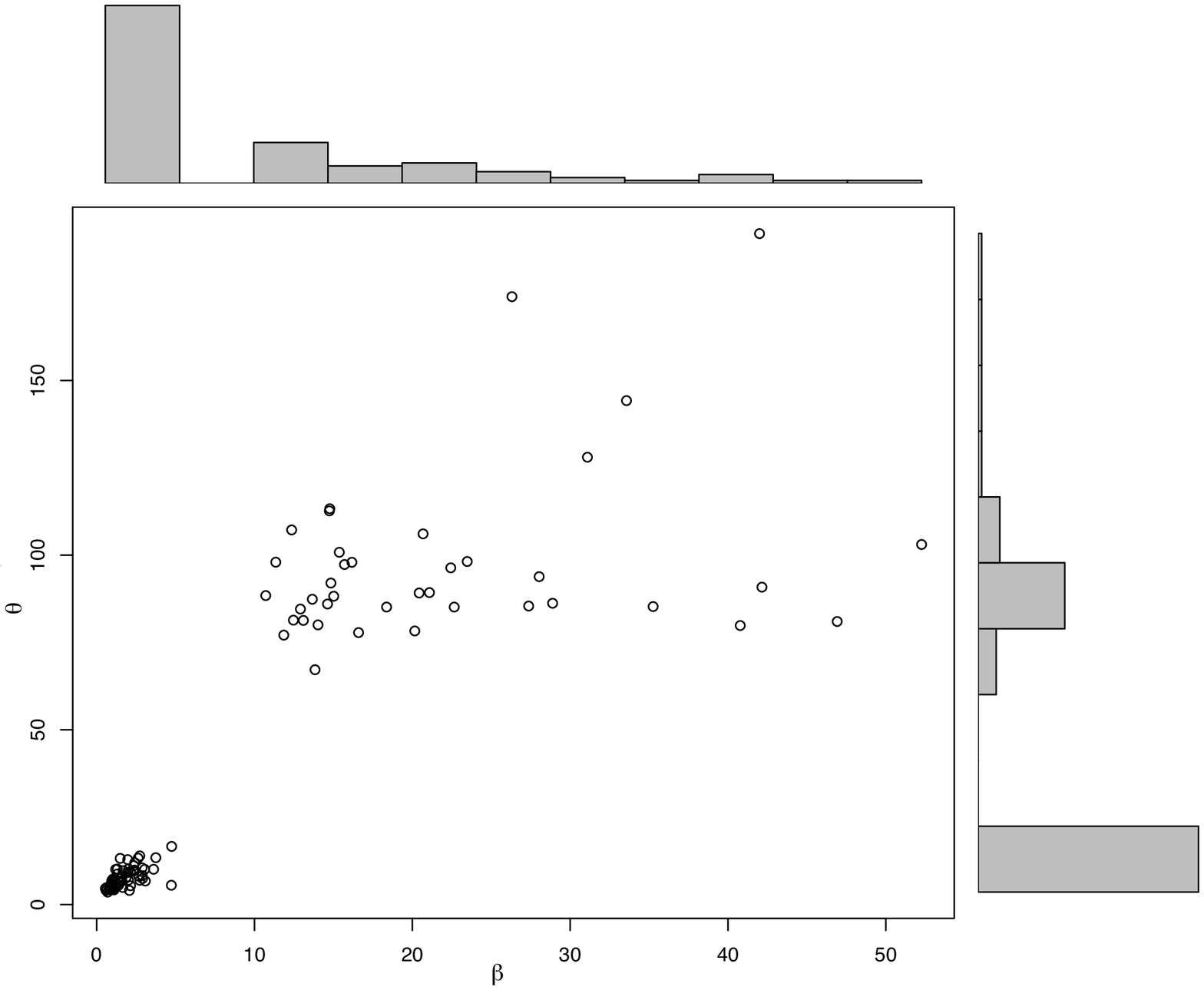}
\begin{figure}[!htp]
\caption*{{\bf Figure S6. Relationship between inferred
    values of parameters in phylogenetic reconstruction of the
    evolutionary history of yeast pathway regulation, under an
    Ornstein-Uhlenbeck model.}  In the main plot, each data point
  reports the results of inference of the evolutionary history of
  regulation of a yeast pathway of size 100: expression data were
  simulated under an Ornstein-Uhlenbeck (OU) model in which the rates
  of regulatory evolution of pathway genes were drawn from an
  inverse-gamma distribution with $\alpha = 3$ and $\beta = 2$ and the
  OU constraint parameter $\theta$ was set to 10, after which
  parameter values for an OU model were optimized against the
  simulated expression data.  For histograms at top and left, the
  independent variable is shared with the axis of the main plot and
  reports the indicated parameter value, and the dependent variable
  reports the proportion of simulated data sets in which the
  corresponding value was inferred.  Note that inferences from most
  simulated data sets accurately estimate $\beta$ and $\theta$, but
  for a few data sets, large parameter values are inferred.}
\end{figure}

\newpage

\includegraphics[width=.8\textwidth]{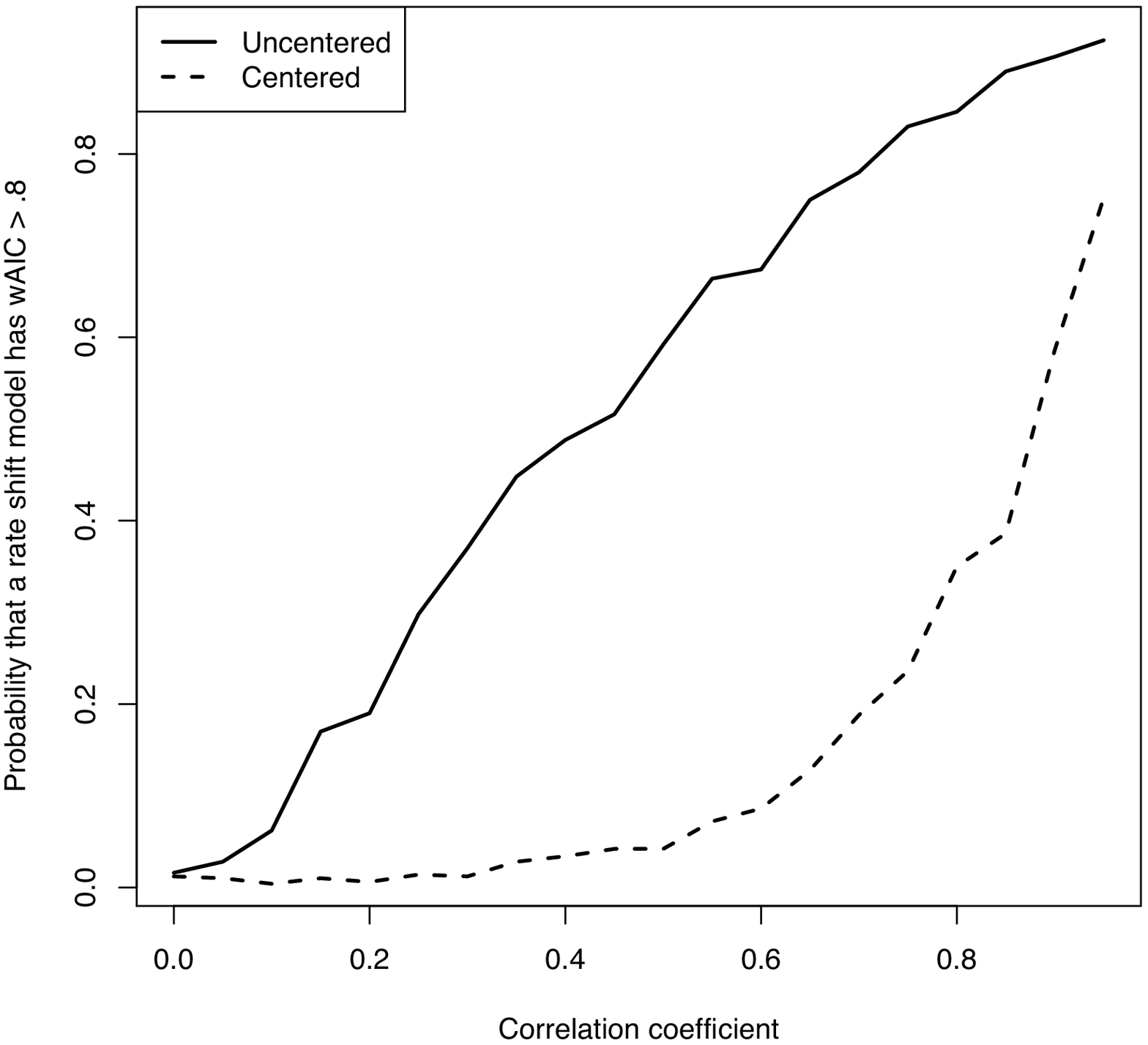}
\begin{figure}[!htp]
\caption*{{\bf Figure S7. Mean-centering pathway
    expression levels in each species corrects for spurious inference
    of non-neutral regulatory evolution arising from gene
    co-regulation.}  Each trace reports the results of inference of
  the evolutionary history of regulation of a yeast pathway of size
  100, from expression data simulated under a Brownian motion model in
  which evolutionary rates were the same on all branches of the yeast
  phylogeny, and pathway genes were correlated with one another with
  respect to expression throughout the phylogeny.  Each line style
  reports one scheme for normalization of simulated expression data
  before evolutionary inference: expression measurements were analyzed
  as is (Uncentered), or the distribution of expression across pathway
  genes for each species in turn was normalized to have a mean of $0$
  (Centered).  The $x$ axis reports the value of the correlation
  coefficient between genes in the group, and the $y$ axis reports the
  fraction of 500 simulations that resulted in a model other than the
  Brownian motion equal-rates model having an Akaike weight greater
  than 0.8.}
\end{figure}

\newpage

\includegraphics[width=.8\textwidth]{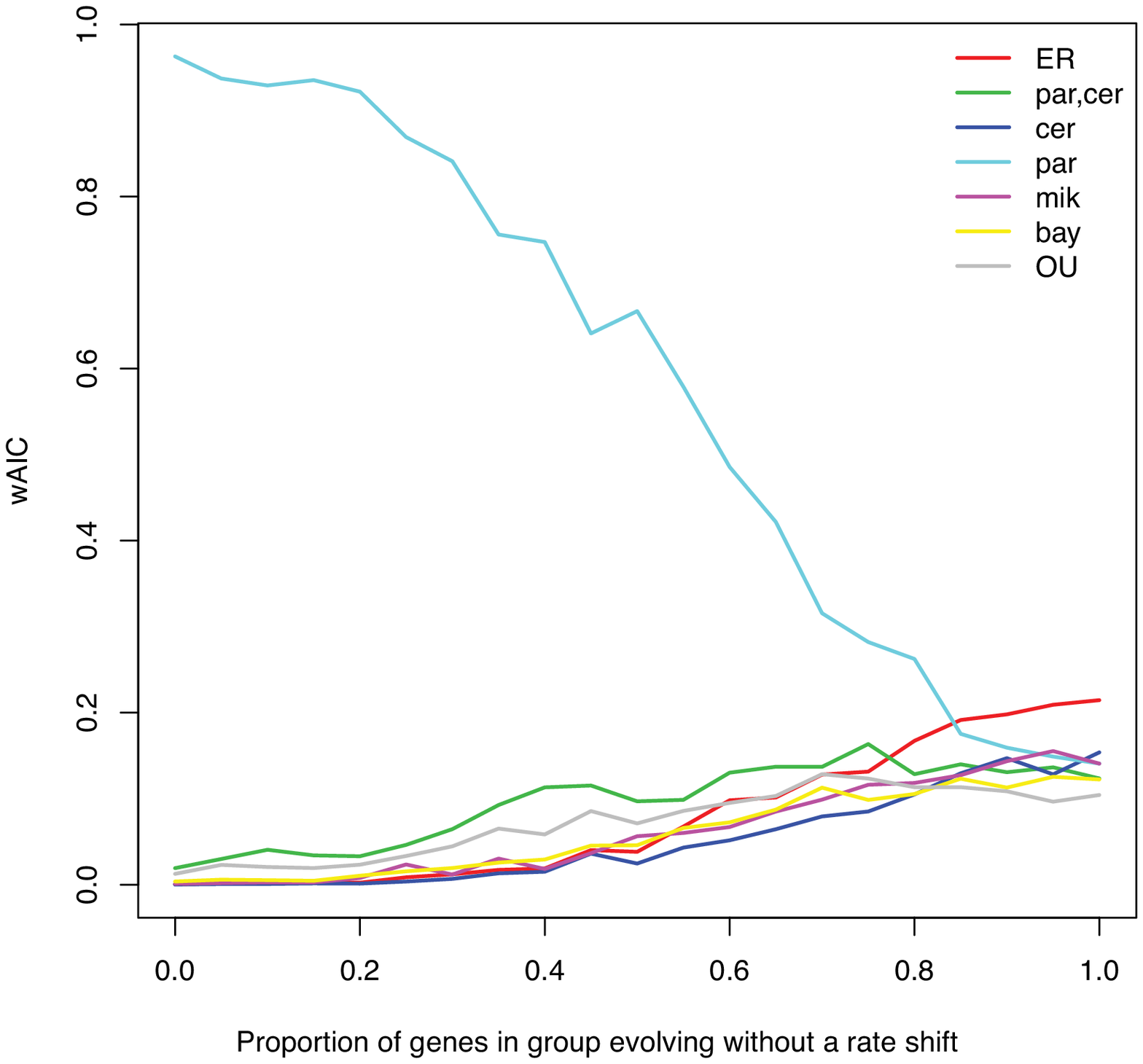}
\begin{figure}[!htp]
\caption*{{\bf Figure S8. Heterogeneity in the mode of
    regulatory evolution across the genes of a pathway has little
    impact on inference of evolutionary histories from expression
    data.}  Each trace reports the results of inference of the
  evolutionary history of regulation of a yeast pathway of size 100,
  from expression data simulated under a Brownian motion model in
  which the rate of regulatory evolution for each gene was drawn from
  an inverse-gamma distribution with $\alpha = 3$, $\beta = 2$ and,
  for the branch leading to \emph{S. paradoxus}, increased by a factor
  of $5$ for a subset of pathway genes.  The $x$ axis reports the
  fraction of genes in the group without a rate shift, and the $y$
  axis reports the average Akaike weight assigned to each model.  Line
  styles are as in Figure 1a of the main text.}
\end{figure}

\newpage

\section*{Supplementary Table Legends}

\begin{table}[!htp]
\caption*{
{\bf Table S1. Strains used in this work.} 
}
\end{table}

\begin{table}[!htp]
\caption*{ {\bf Table S2. Read mapping statistics from
    yeast RNA-seq.} Each set of rows reports the mapping statistics
  for reads from RNA-seq libraries used for a comparison of two yeast
  species.  For a given set, in row headings, numerals indicate
  biological replicates, single species names indicate homozygotes,
  and species name pairs separated by a slash indicate diploid
  interspecies hybrids.  Each row reports results from one library.
  Total reads, the full set of reads sequenced.  Have polyT, the
  number of reads containing at least two consecutive Ts at only one
  end.  Uniquely mapped, the number of reads mapping uniquely, with no
  mismatches, to the concatenated genomes of the two species of the
  set.  Passed through filters, the number of reads whose poly-A tails
  were unlikely to have originated from oligo-dT mispriming to A-rich
  regions of the genome; see Methods.}
\end{table}

\begin{table}[!htp]
\caption*{
 {\bf Table S3.  Fitted models of $cis$-regulatory evolution in yeast pathways.} Data are as in Table 1 of the main text except that results for all pathways are shown.
}
\end{table}

\begin{table}[!htp]
\caption*{
  {\bf Table S4.  Fitted models of species regulatory evolution in yeast pathways.}  Data are as in Table 2 of the main text except that results for all pathways are shown.
}
\end{table}

\end{document}